\documentclass[journal]{IEEEtran}

\usepackage{xcolor,soul,framed} 
\usepackage[colorlinks]{hyperref}
\usepackage{algorithm}
\usepackage{algorithmic}
\usepackage{subfigure}
\usepackage{url}
\usepackage{cite}
\colorlet{shadecolor}{yellow}
\usepackage[pdftex]{graphicx}
\usepackage{graphicx}
\DeclareGraphicsExtensions{.pdf,.jpeg,.png}

\usepackage{amsmath,amssymb,amsfonts}
\usepackage{array}
\usepackage{mdwmath}
\usepackage{mdwtab}
\usepackage{eqparbox}
\usepackage{url}
\IEEEoverridecommandlockouts
\usepackage{soul}
\usepackage{threeparttable}

\begin{document}
\title{On Propagation Characteristics of Reconfigurable Surface Wave Platform: Simulation and Experimental Verification}
\author{Zhiyuan~Chu,~\IEEEmembership{Student Member,~IEEE,} 
Kin-Fai~Tong,~\IEEEmembership{Fellow,~IEEE,} 
Kai-Kit~Wong,~\IEEEmembership{Fellow,~IEEE,}\\ 
Chan-Byoung Chae,~\IEEEmembership{Fellow,~IEEE,} 
Chi Hou Chan,~\IEEEmembership{Fellow,~IEEE,} 

\thanks{The work of Z. Chu, K. K. Wong and K. F. Tong is supported by the Engineering and Physical Sciences Research Council (EPSRC) under Grant EP/V052942/1. For the purpose of open access, the authors will apply a Creative Commons Attribution (CC BY) licence to any Author Accepted Manuscript version arising.}
\thanks{The work of C. B. Chae is supported by the Institute of Information and Communication Technology Promotion (IITP) grant funded by the Ministry of Science and ICT (MSIT), Korea (No. 2021-0-02208, No. 2021-0-00486).}
\thanks{The work of C. H. Chan is supported by the State Key Laboratory of Terahertz and Millimeter Waves, City University of Hong Kong, Hong Kong.}
\thanks{Z. Chu, K. K. Wong, and K. F. Tong are with the Department of Electronic and Electrical Engineering, University College London, Torrington Place, WC1E 7JE, United Kingdom.}
\thanks{K. F. Tong and C. H. Chan are with the State Key Laboratory of Terahertz and Millimeter Waves, City University of Hong Kong, Hong Kong.}
\thanks{K. K. Wong and C. B. Chae are with School of Integrated Technology, Yonsei University, Seoul, 03722, Korea.}
\thanks{{\em Corresponding author:} K. F. Tong (e-mail: $\rm k.tong@ucl.ac.uk$).}
}

\markboth{Submitted to IEEE TRANSACTIONS ON ANTENNAS AND PROPAGATION, 2023
}{ \MakeLowercase{\textit{}}: }

\maketitle

\begin{abstract}
Reconfigurable intelligent surface (RIS) as a smart reflector is revolutionizing research for next-generation wireless communications. Complementing this is a concept of using RIS as an efficient propagation medium for potentially superior path loss characteristics. Motivated by a recent porous surface architecture that facilitates reconfigurable pathways with cavities filled with fluid metal, this paper studies the propagation characteristics of different pathway configurations in different lossy dielectric materials on the reconfigurable surface wave platform by using a commercial full electromagnetic simulation software and S-parameters measurements. This paper also looks into the best scheme to switch between a straight pathway and a $90^\circ$-bend and attempts to quantify the additional path loss when making a $90^\circ$ turn. The experimental results verify the simulation results, showing the effectiveness of the proposed reconfigurable surface wave platform for a wide-band, low path loss and highly programmable communications. 
\end{abstract}


%
\IEEEpeerreviewmaketitle

\section{Introduction}
\IEEEPARstart{W}{ith} research blossoming for the sixth-generation (6G) mobile communications, a few white papers have already identified reconfigurable intelligent surface (RIS) as one of the key enabling technologies, e.g., \cite{IMT-2030-white,5gic-white}. RIS is enthusiastically motivated by its relatively low-cost deployment and great coverage performance \cite{di2021catching}. Recently, there is a twist by extending RIS to serve partly as a programmable propagation medium for less path loss and more controlled communication and partly as an intelligent reflect-beamformer \cite{shojaeifard2022mimo, wong2020vision}. 

Compared to the propagation of traditional space waves used in mobile communications, surface waves propagate on a 2-dimensional (2D) dielectric-coated conductor plate with a path loss which is proportional to propagation distance $d$ rather than the less desirable square of $d$ in the case of space waves \cite{sarkar2017surface,cullen1954excitation}. This indicates that the use of surface waves might provide better propagation conditions, e.g., stronger desired signals and less interference, for mobile communications \cite{shojaeifard2022mimo,wong2020vision}. In recent years, it is encouraging to witness that surface waves also find applications in networks-on-chip systems \cite{karkar2022thermal,fuschini2018ray,schafer2018chip}, wearable devices for on-body networks \cite{berkelmann2021antenna,turner2012novel} and replacing cables to reduce through-life costs in industrial environments \cite{turner2013surface}.

The propagation of surface waves is typically characterized by a non-directional wavefront emanating from the transmitting terminal and then spreading onto the entire plane. Previous studies investigated how to manipulate surface waves using techniques such as checkerboard surface \cite{gonzalez2014surface}, leaky-wave concept \cite{kuznetcov2019printed}, holographic wide-band beam-scanning \cite{moeini2019wide} and printed planar structures in specific geometries \cite{tcvetkova2020perfect}. Isotropic and anisotropic surface wave cloaking techniques are further proposed in \cite{mcmanus2016isotropic}, and anisotropic materials are used to reduce scattering loss when surface waves propagate past a sharp angle at $6\,{\rm GHz}$ in \cite{xu2015broadband}. Transformation optics \cite{yang2012accurate} and stacked Eaton lenses \cite{wei2020surface} are also found to be useful in directing surface waves into intended directions via refraction in the terahertz band. 

\begin{figure}[] 
\centering 
\includegraphics[width=0.9\columnwidth]{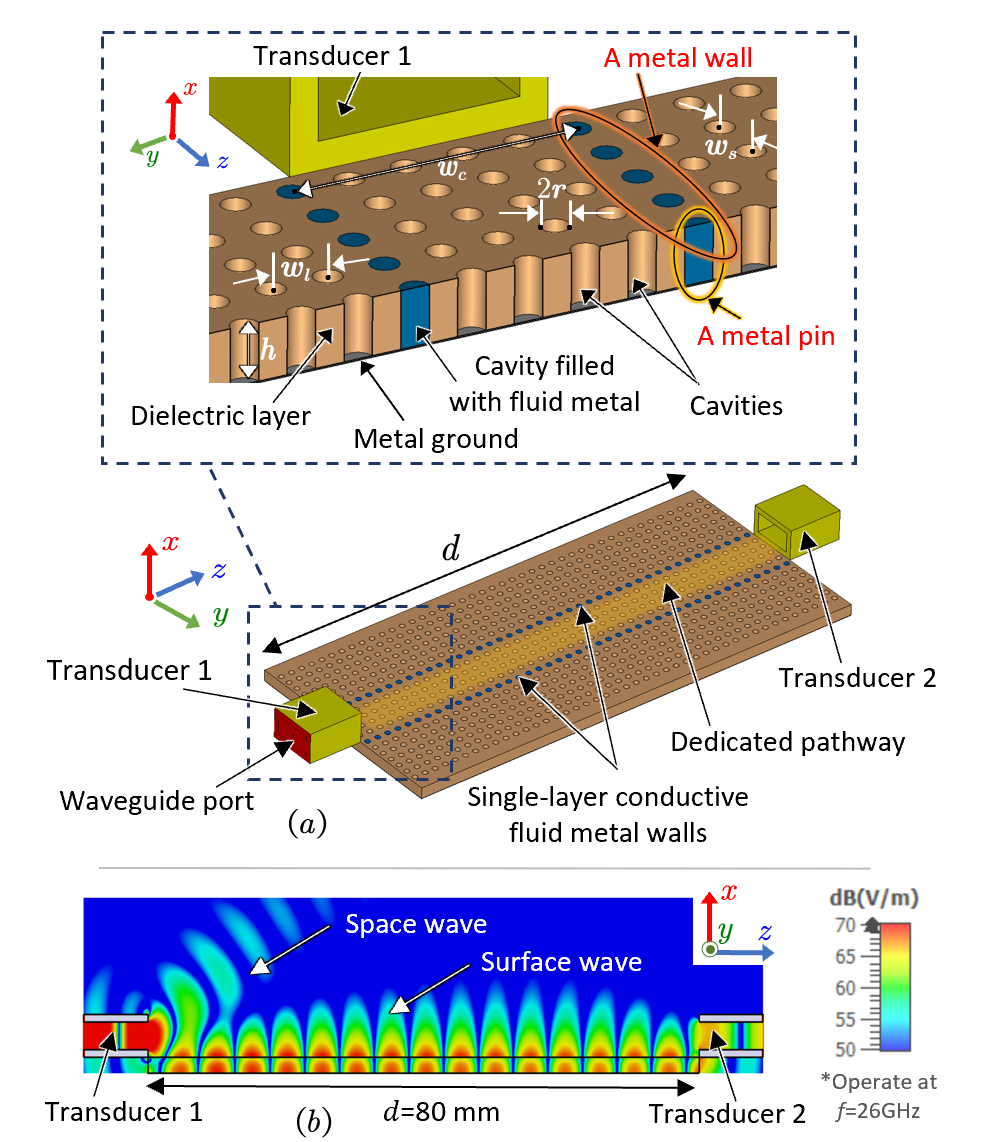} 
\vspace{-0.5cm}
\caption{$ (a)  $A reconfigurable surface with evenly distributed cavities that can be filled with conductive fluid metal and $ (b)  $ its simulation result.}\label{Basic_structure} 
\vspace{-0.5cm}
\end{figure}

However, there has not been much progress in developing reconfigurable hardware architectures that could dynamically control the propagation direction of surface waves. Motivated by this, the objective of this paper is to devise a reconfigurable surface wave platform that can create guided pathways on demand, and evaluate such design using 3-dimensional (3D) electromagnetic simulations and experiments. The design can utilize either fluid metal, such as Galinstan, or solid metallic conducting pins to isolate and guide surface waves into specific pathways. The selection of fluid metal is highly motivated by its high conductivity and low adhesion \cite{hensel2014fluid} and fluid metal has already been applied in many fields, such as microfluidics \cite{battat2022nonlinear} and liquid metal antennas \cite{liu2022liquid}. The proposed reconfigurable surface is based on a porous surface model where small cylindrical cavities are evenly distributed over a 2D platform as shown in Fig.~\ref{Basic_structure}. The working mechanism is simple and effective.

Conductive fluid metal can be filled into the selected cylindrical cavities by using micro-pumps under the metal ground plane to create a surface wave propagation pathway. Low adhesion and high fluidity fluid metal can be pumped in or out of the cavities through the microtubes attached to the bottom side of the porous dielectric layer \cite{tang2014liquid}. In so doing, the propagation direction of surface waves can be predictably guided in the dedicated pathway created by the fluid metal columns and can be altered by changing the pathway formation. 

The proposed porosity-based reconfigurable surface wave platform is first considered in \cite{chu2022surface}. However, the emphasis of \cite{chu2022surface} lies on the performance of different porosity density and pattern, only simulation results were presented. However, the contributions of this paper are twofold. First of all, we report experimental results of a 3D-printed surface prototype and compare them with full 3D electromagnetic simulation results. Secondly, in this paper, we investigate how a $90^\circ$-bend can be designed and quantify the incurred insertion loss. More specifically, we will examine the following characteristics:
\begin{itemize}
\item The surface wave path loss of a straight guided pathway in dielectric materials of different loss tangents in the millimeter-wave band.
\item The path loss and isolation of different guided pathways formed by single-layer or multi-layer metal pin walls.
\item The frequency dependency of the path width.
\item The level of reconfigurability between a straight pathway and $90^\circ$-bend.
\item The power attenuation of different corner configurations in a $90^\circ$-bend.
\end{itemize}
In summary, our results contribute to understanding the propagation characteristics of surface waves in physical environments and offer experimental evidence on the feasibility of the porosity-based reconfigurable surface wave platform.

\section{Porosity-Based Reconfigurable Surface} 
\subsection{Geometry} 
As shown in Fig.~\ref{Basic_structure}$ (a) $, we consider a porous reconfigurable surface geometry consisting of a top porous dielectric layer sitting on a conductive metal ground. Surface wave is excited by Transducer $ 1 $, in the form of a rectangular waveguide, and then propagates along the interface between the top surface and air in an open environment to Transducer $ 2 $ in the $ +z $-direction on the right end. As shown in Fig.~\ref{Basic_structure}$(b)$, we can see that most of the wave is bound by the dielectric-coated metal surface as surface wave. To maximize the surface wave excitation efficiency, an optimized launcher can be used \cite{wan2019simulation}. Surface waves can be considered as a specific solution of the cylindrical wave in the Transverse Magnetic (TM) mode, given by \cite{Barlow1958surface}
\begin{align}
H_y&=\frac{A}{\sqrt{d}}e^{-\gamma _zz}e^{-\gamma _xx}e^{j\omega t},\label{H_y}\\
E_x&=\frac{A\gamma _z}{j\omega \varepsilon _0\sqrt{d}}e^{-\gamma _zz}e^{-\gamma _xx}e^{j\omega t},\label{E_x}\\
E_z&=-\frac{A\gamma _x}{j\omega \varepsilon _0\sqrt{d}}e^{-\gamma _zz}e^{-\gamma _xx}e^{j\omega t},\label{E_z}
\end{align}
where $A$ is the amplitude, $d$ denotes the surface wave propagation distance along the surface, $\omega $ is the angular frequency, $\varepsilon _0=8.854\times 10^{-12}\,{\rm F/m}$ is the vacuum permittivity, $\gamma _x$ and $\gamma _z$ are the propagation coefficients vertically away from and horizontally along the surface in the $+x$ and $+z$-direction, respectively.

For the proposed porous reconfigurable surface in which the cylindrical cavities are evenly distributed inside the dielectric layer, as depicted in Fig.~\ref{Basic_structure}$ (a) $, the relative permittivity $ \varepsilon_r$ of a solid dielectric layer becomes the effective relative permittivity $\varepsilon_r^{\rm eff}$ which is dependent on the surface porosity 
\begin{equation}\label{eq:rho}
\rho=\frac{S_{\rm cavity}}{S_{\rm surface}},
\end{equation}
where $S_{\rm cavity}$ is the circular top surface area of the cavities and $S_{\rm surface}$ is the total area of the top dielectric layer surface. In \cite{chu2022surface}, it has been illustrated that a surface with high porosity performs similar to a homogeneous solid surface. Specifically, $\varepsilon_r^{\rm eff}$ can be obtained by \cite{liu2016general} 
\begin{equation}
\varepsilon_r^{\rm eff}=\frac{\varepsilon_r\left[1+3\varepsilon_r+3\rho(1-\varepsilon_r)\right]}{1+3\varepsilon_r +\rho(\varepsilon_r-1)}.
\end{equation}
\subsection{Working Mechanism} 

\begin{figure}[]
\centering
\includegraphics[width=0.9\columnwidth]{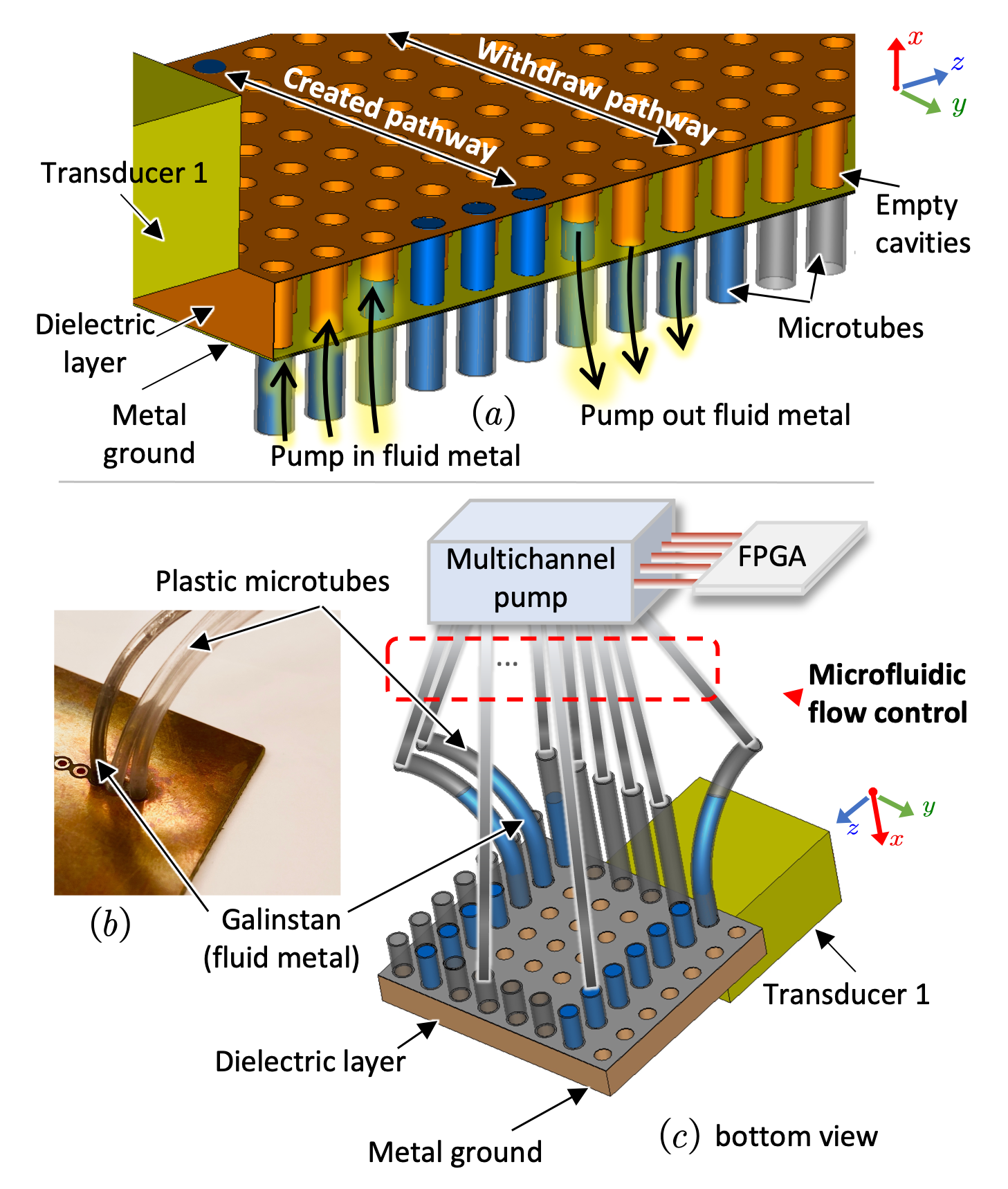}
\caption{Illustration of the working mechanism of the proposed reconfigurable surface wave platform: $ (a)  $ The fluid metal is pumped in or out of the cavities from the microtubes attached to the bottom ground plane to create or withdraw the surface wave pathway. $ (b) $ shows a surface wave platform prototype using Galinstan. And $  (c)  $ illustrate the combined operation of the reconfigurable surface and micro-fluidics by the mean of a multichannel pump and FPGA board controlling the flow of fluid metal.}\label{Basic_pump}
\vspace{-0.5cm}
\end{figure}

The working mechanism is shown in Fig.~\ref{Basic_pump}$(a)$, conductive fluid metal is introduced into desired cavities by the micro-pumps under the metal ground plane to form a dedicated surface wave pathway. And the fluid metal can be pumped in or out of the cavities from the microtubes attached to bottom to create or withdraw the pathway on demand as depicted in the figure. A prototype surface using Galinstan, a conductive fluid alloy, is shown in Fig.~\ref{Basic_pump}$(b)$. It is also possible to create different surface wave propagation pathway by arranging fluid metal pins in columns to assemble a `fluid metal wall' \footnote{In this paper, the word `wall' is used to represent a row of cavities filled with fluid metal even though it is not a solid structure standing above the surface.}.  Furthermore, as shown in Fig.~\ref{Basic_pump}$(c)$, the reconfigurable surface integrated with fluid metal is expected to be operated in combination with a multichannel programmable pump, such as peristaltic pump \cite{peristaltic}. By applying microfluidics technology, precise manipulation and control of small volumes of fluid metal at the microscale can be achieved. A FPGA board is connected to the multichannel programmable pump for determining the optimum pathways for specific surface wave propagation.

\begin{figure}[]
\centering
\includegraphics[width=0.9\columnwidth]{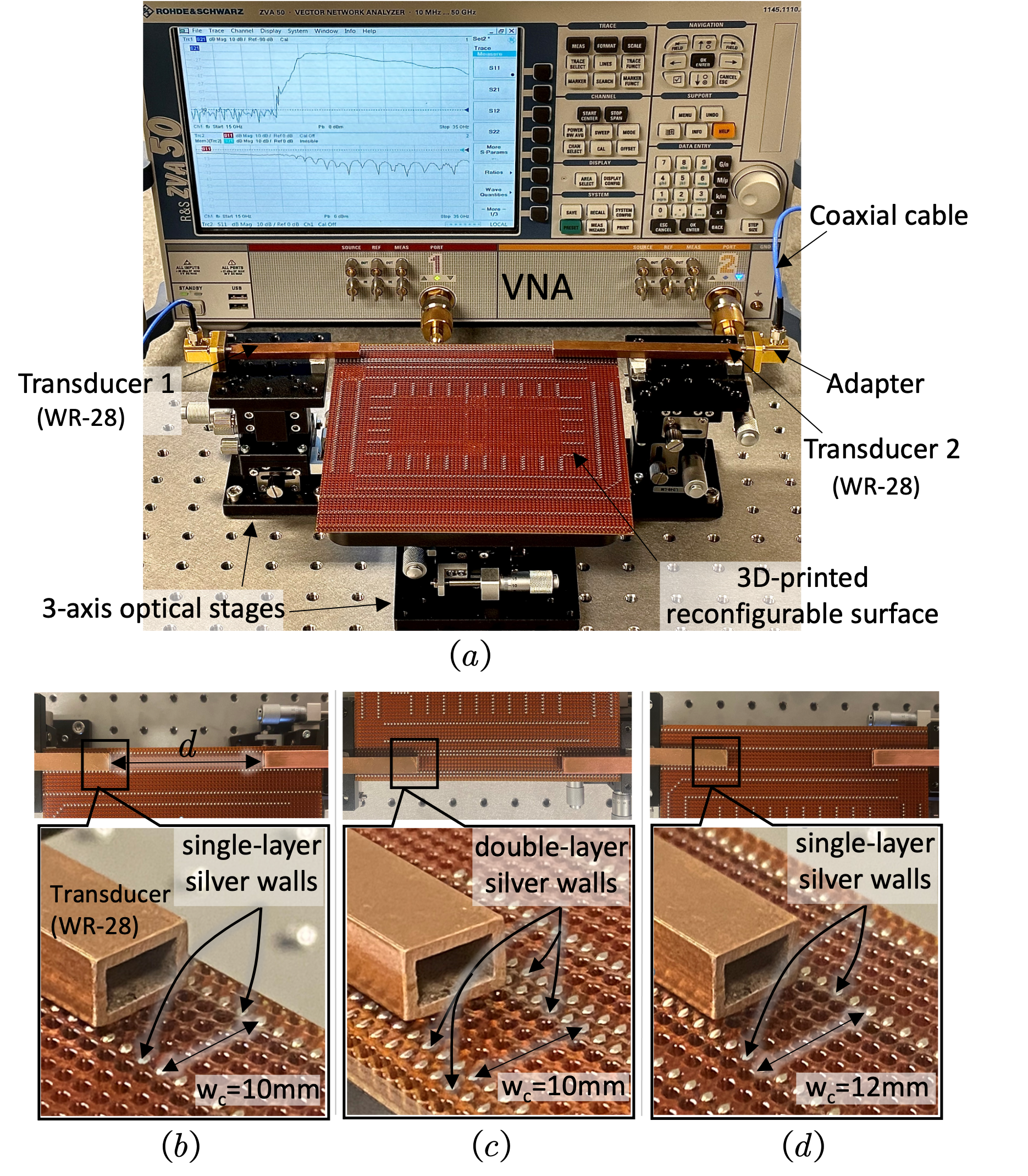}
\vspace{-0.5cm}
\caption{The measurement setup for $(a)$ a 3D-printed reconfigurable surface prototype where the straight surface wave pathway is formed by $(b)$ single-layer silver pin walls or $(c)$ double-layer silver pin walls with a pathway width of $10\,{\rm mm}$ or $(d)$ single-layer silver pin walls with a pathway width of $12\,{\rm mm}$.}\label{Basic_3D_prototype}
\vspace{-0.5cm}
\end{figure}

\begin{table}[]
\caption{Key parameters in the measurements and simulations}\label{Table1}
\vspace{-1.0mm}
\begin{tabular}{llr} 
\cline{1-2} 
 \cline{1-2} 
{Parameter} & {Value}\\
 \cline{1-2} 

\mbox{operating frequency}, $f$ & $26, 28, 30\,{\rm GHz}$\\
\mbox{transducer (WR-28) frequency band}, $f_b$ & $21-42\,{\rm GHz}$\\
\mbox{height of transducer aperture}, $h_a$ & $3.566\,{\rm mm}$\\
\mbox{width of transducer aperture}, $w_a$ & $7.112\,{\rm mm}$\\
\mbox{thickness of transducer}, $l_a$ & $1\,{\rm mm}$\\
\mbox{WR-28 to 2.92mm coaxial adaptor}, & $26.5-40\,{\rm GHz}$\\
\mbox{} & VSWR: 1.2:1 (Typ.)\\
\mbox{relative permittivity of PTFE layer}, $\varepsilon_r$ & $2.1$ \\
\mbox{effective permittivity of PTFE layer}, $\varepsilon _r^{\mathrm{eff}}$ & $1.86$\\
\mbox{loss tangent} of low-loss PTFE, $\tan\delta @ 26\,{\rm GHz}$ & $0.00005$ \\
\mbox{loss tangent} of PTFE, $\tan\delta @ 26\,{\rm GHz}$ & $0.006$\\
\mbox{thickness of PTFE layer}, $h$ & $3\,{\rm mm}$\\
\mbox{relative permittivity of 3D-print resin layer}, $\varepsilon_r$ & $2.8$ \\
\mbox{effective permittivity of 3D-print resin layer}, $\varepsilon _r^{\mathrm{eff}}$ & $2.4$\\
\mbox{loss tangent} of 3D-print resin, $\tan\delta @ 26\,{\rm GHz}$ & $0.0155$\\
\mbox{thickness of 3D-print resin layer}, $h$ & $2\,{\rm mm}$\\
\mbox{pathway width}, $w_c$ & $10,12\,{\rm mm}$\\
\mbox{radius of cavity}, $r$ & $0.5\,{\rm mm}$\\
\mbox{center-to-center separation between cavity}, $w_s$ & $2\,{\rm mm}$\\
\mbox{surface porosity}, $\rho$ & $19.63\%$\\
\mbox{propagation distance}, $d$ & $50-150\,{\rm mm}$\\
\mbox{thickness of the metal ground}, $h_m$ & $0.05\,{\rm mm}$\\
\mbox{electrical conductivity of silver ink}, $\sigma_s$ & $3.15 \times 10^6~\,{\rm S/m}$\\
\mbox{electrical conductivity of Galinstan}, $\sigma_g$ & $3.46 \times 10^6~\,{\rm S/m}$\\
\mbox{vacuum permittivity}, $\varepsilon_0$ & $8.854\times 10^{-12}~\,{\rm F/m}$\\
\mbox{surface impedance in the freq. band}, $Z_s$ & $ j240 - j330 {\rm \Omega}$ \\
\cline{1-2} 
\cline{1-2} 
\end{tabular}
\vspace{-1.5mm}
\end{table}

\section{Results and Discussions} 
\subsection{Experimental Setup}

\begin{table*}[]
\caption{Comparison of different transmission lines}\label{Table2}
\vspace{-3mm}
\centering
\begin{tabular}{ccccc}
\hline
Transmission mode                                                  & \begin{tabular}[c]{@{}c@{}}Substrate material /\\Type\end{tabular}   & \begin{tabular}[c]{@{}c@{}}Working frequency, $ f $\\ (${\rm GHz}$)\end{tabular}  & \begin{tabular}[c]{@{}c@{}}Loss tangent, $\tan\delta$\end{tabular} & \begin{tabular}[c]{@{}c@{}}Path loss, $ L $\\ (${\rm dB/m}$)\end{tabular} \\ \hline
\begin{tabular}[c]{@{}c@{}}Reconfigurable surface wave pathway\\ \end{tabular} & Low-loss PTFE \cite{PTFE2015}& 26 & 0.00005 & 0.84 \\ \hline
\begin{tabular}[c]{@{}c@{}}Reconfigurable surface wave pathway\\ \end{tabular}& PTFE & 26 & 0.006 & 14.5 \\ \hline
\begin{tabular}[c]{@{}c@{}}Reconfigurable surface wave pathway\\ \end{tabular} &3D-print resin \cite{printed2022} & 26 & 0.0155 & 28.1 \\ \hline
\begin{tabular}[c]{@{}c@{}}Reconfigurable transmission line \cite{gonzalez2014basic}\\\end{tabular}&Roger 6010 & 26 & 0.0023 & 14.3 \\ \hline
\begin{tabular}[c]{@{}c@{}}Controllable mode transmission line \cite{zhang2017differential}\\\end{tabular}&Roger 4003 & 4 & 0.0027 & 6.25 \\ \hline
\begin{tabular}[c]{@{}c@{}}Microstrip transmission line \cite{iPCB}\\\end{tabular}&Roger 4350B & 24 & 0.0037 & 17.1 \\ \hline
\begin{tabular}[c]{@{}c@{}}Millimetre-wave coaxial cable \cite{Huber_Suhner}\end{tabular}& Sucoflex-103& 26 & — & 1.65 \\ \hline
\end{tabular}
\end{table*}

Fig.~\ref{Basic_3D_prototype}$(a)$ illustrates the experimental setup of the 3D-printed reconfigurable surface platform prototype. The center-to-center separation between two silver columns on the porous platform is defined as the pathway width $w_c$ of the straight pathway. To match the width of the pathway, two commercial WR-28 rectangular waveguides with a wall thickness of $1\,{\rm mm}$ are selected as the transducers. They are located at either end of the pathway and secured on the 3-axis optical linear stages which positions can be adjusted precisely for different propagation distances $d$. The ${\rm S}_{21}$ parameter is measured by the vector network analyzer (VNA) connected to the transducers through two coaxial-to-waveguide adaptors and high frequency coaxial cables. It is observed that the approximately $0.5\,{\rm mm}$ air gap between the bottom of transducer and the top of surface, and the $1\,{\rm mm}$ thick waveguide wall have noticeable impacts on the measurement. Such issue can be effectively addressed by incorporating a specially designed transducers with impedance matching to achieve tight coupling to the surface as described in our previous work \cite{wan2019simulation}. Therefore, we focus more on the propagation characteristics of surface waves on the reconfigurable platform rather than matching the interfaces in this paper. To eliminate the impacts, the path loss and insertion loss results presented in the paper have been processed by subtracting the ${\rm S}_{21}$ between different pathways, such as different path lengths or different path configurations, rather than directly using the measured ${\rm S}_{21}$ of the pathways.  The full physical dimensions of the reconfigurable surface are presented in TABLE \ref{Table1}.

\begin{figure}[]
\centering
\includegraphics[width=0.9\columnwidth]{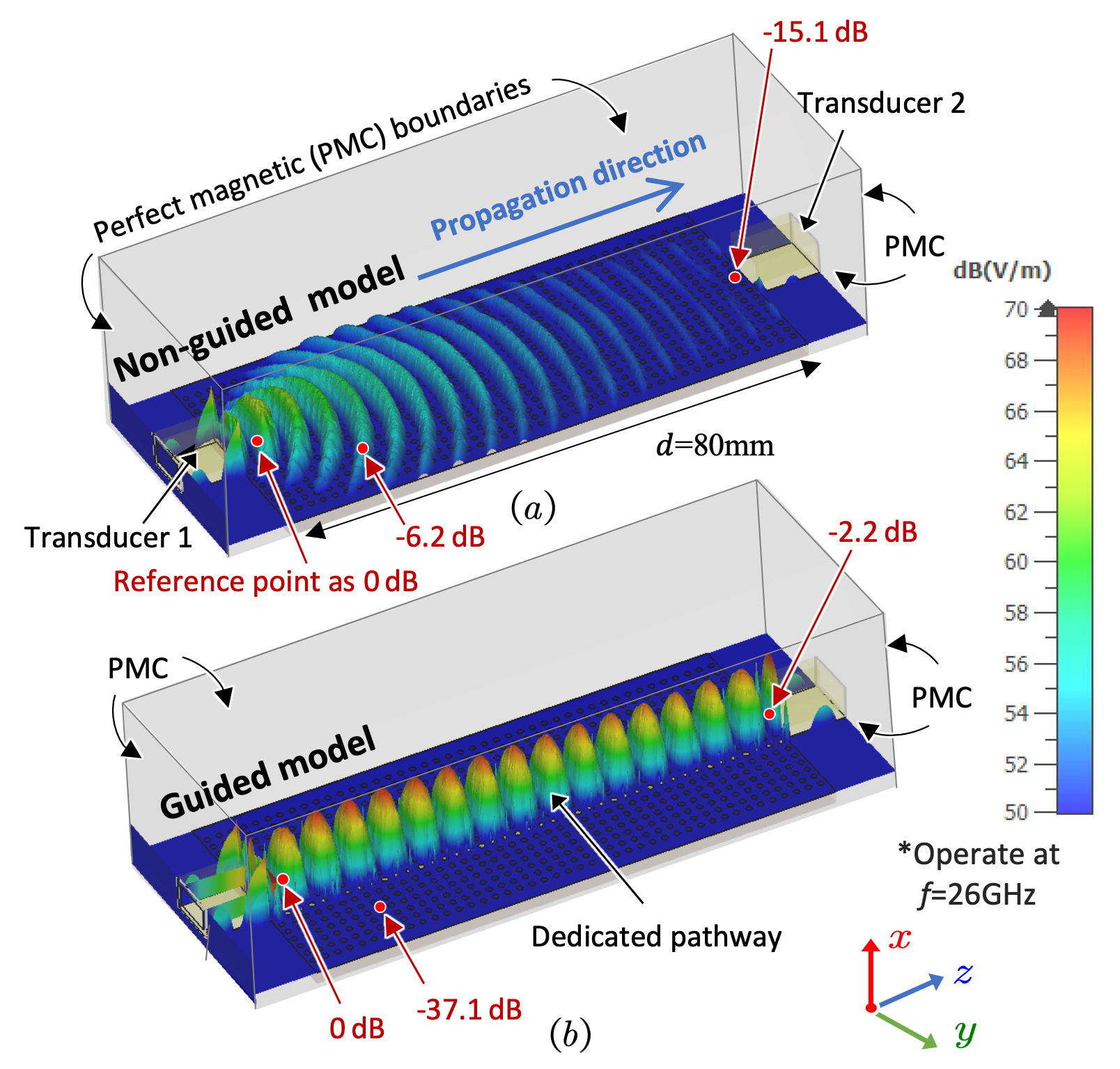}
\vspace{-0.2cm}
\caption{The electric field contour of $(a)$ a non-guided surface, illustrating that the surface wave propagates over the entire surface, and $(b)$ a guided model where the surface wave is guided along a straight dedicated pathway by two columns of fluid metal walls in the reconfigurable surface geometry.}\label{Basic_CST}
\vspace{-0.2cm}
\end{figure}

\begin{figure}[]
\centering
\includegraphics[width=0.9\columnwidth]{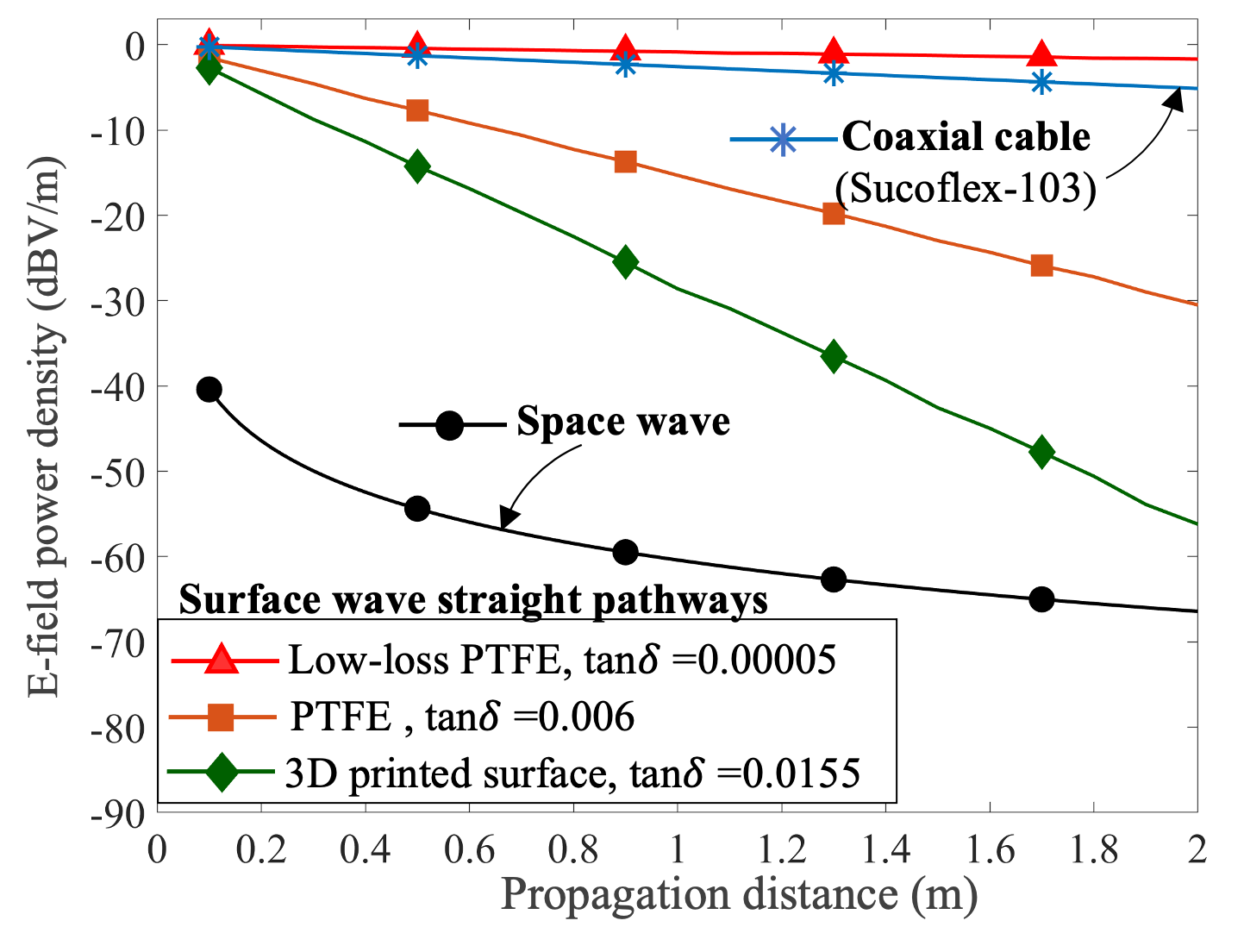}
\vspace{-0.2cm}
\caption{Simulated E-field power density $\rm (dB\,V/m)$ for a $2m$ straight pathway created by Galinstan (fluid metal) walls, with different dielectric surfaces: low-loss PTFE ($\tan\delta=0.00005$), PTFE ($\tan\delta=0.006$), and 3D print resin ($\tan\delta=0.0155$). A coaxial cable (Sucoflex-103) is also included for comparison.}\label{PTFE}
\vspace{-0.5cm}
\end{figure}

\begin{figure}[]
\centering
\includegraphics[width=0.9\columnwidth]{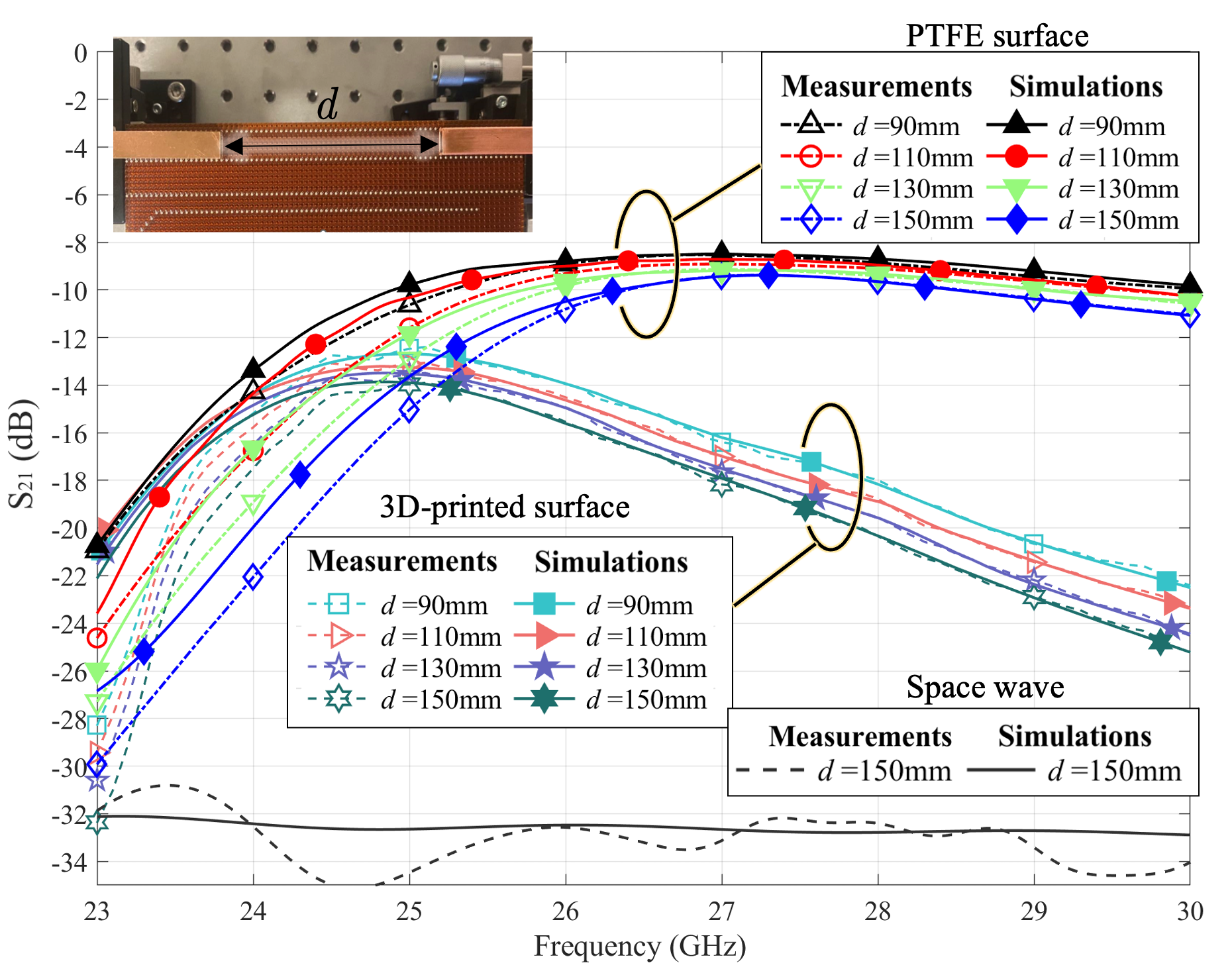}
\vspace{-0.5cm}
\caption{${\rm S}_{21}$ results for measurement and simulation inside the straight pathway for various propagation distance $d$ from $50\,{\rm mm}$ to $150\,{\rm mm}$.}\label{single_layer_result1}
\vspace{-0.5cm}
\end{figure}

\subsection{Simulation Results}

The 3D electromagnetic simulation results in Fig.~\ref{Basic_CST}$(b)$ demonstrate that surface wave can be guided within a straight pathway by two columns of fluid metal walls on the proposed reconfigurable surface. Contrarily, Fig.~\ref{Basic_CST}$(a)$ shows that surface wave spreads over the entire surface in the non-guided geometry where the cavities are void. To provide a fair comparison, we normalize the electric field power density on the surface by the corresponding value at the reference point (marked in Fig.~\ref{Basic_CST}$(a)$ and $(b)$ ), which is $5\,{\rm mm}$ in front of the aperture of Transducer $ 1 $. The results indicate that the surface wave decays to $-2.2\,{\rm dB}$ at the aperture of Transducer $ 2 $  on the straight pathway with a $80\,{\rm mm}$ propagation distance while it drops to $-15.1\,{\rm dB}$ in the non-guided case. It is evident that the E-field is significantly concentrated within the straight pathway. Moreover, the power outside the pathway is kept to be as low as $ -37.1\,{\rm dB} $, while it is $-6.2\,{\rm dB}$ at the same position in the non-guided case, indicating that the proposed reconfigurable surface can create dedicated and isolated pathways for efficient surface wave communications.

Fig.~\ref{PTFE} shows the simulated E-field power along the straight guided surface wave pathways which are made of dielectric materials with different loss tangents. The thicknesses of the materials, limited by their availability in the lab, are also chosen for achieving the surface impedance ($Z_s$) near $ j260{\rm \Omega} $ \cite{wan2019simulation}. The length of the pathways are $2\,{\rm  m}$ and the E-field at a distance of $10\,{\rm cm}$ from the aperture of Transducer $ 1 $ is considered as the zero reference for normalization.  Three different dielectric materials: i) low-loss low-density Polytetrafluoroethylene (PTFE) ($\varepsilon_r$ = $2.1$, $\tan\delta=0.00005$, $h=2mm$) \cite{PTFE2015}, (ii) common PTFE ($\varepsilon_r$ = $2.1$, $\tan\delta=0.006$, $h=3mm$), and (iii) 3D-print resin ($\varepsilon_r$ = $2.8$, $\tan\delta=0.0155$, $h = 2mm$) \cite{printed2022} are studied. The path losses in free space and a low-loss RF coaxial cable (Sucoflex-103) are also included in the figure for comparisons. It can be observed that the 3D-print resin surface has a significantly higher propagation loss than that of the PTFEs, however, they are all lower than that in free space. The comparison provides useful reference for selecting or developing materials for applications in different physical scales. In the case of large-scale and multi-user applications, low-loss low-density PTFE dielectric will be a better choice if manufacturing process is not the main concern, as computer numerical control (CNC) machinery may be required in the process, moreover, large panel of low-loss low-density PTFE is yet commercially available. On the other hand, the 3D-print technology can facilitate prompt proof of concept development by printing metal and dielectric simultaneously in this paper. The 3D-printed resin-columnar silver ink is used as the fluid metal, as shown in Fig.~\ref{Basic_3D_prototype}$(b)$--$(d)$, and it is 3D-printed into the dielectric layer and connected to the ground silver layer to save the manufacturing time in the reconfigurable pathway models. The conductivity of the silver ink is about $8.9\%$ lower than that of Galinstain, and it has been verified in the simulation that such difference in conductivity does not significantly affect the reported results. It can be observed that the reconfigurable surface wave platform using low-loss low-density PTFE outperforms the low-loss cable.  Also the significant loss in the 3D-printed platform mainly caused by the lossy resin available in the market at the moment, not the proposed reconfigurable surface wave geometry. Therefore, it will be highly desirable if low-loss 3D-print resin can be available in the near future. The proposed reconfigurable surface wave platform can then support more different kinds of applications when suitable dielectric materials are available. Table \ref{Table2} presents a comparison of path loss of different surfaces and geometries with different dielectric loss tangents, for example, the path losses for a reconfigurable transmission line ($\tan\delta=0.0023$) and a transmission microstrip line ($\tan\delta=0.0037$) are $14.3\,{\rm dB/m}$ and $17.1\,{\rm dB/m}$, respectively. It is clear that the proposed reconfigurable surface wave platforms using common PTFE, which has higher loss tangent, perform better than other 2D re-configurable geometries and it shows its potential for applications in long-distance relay communications, as previously mentioned in our work \cite{shojaeifard2022mimo, wong2020vision}. On the other hand, for compact systems and high precision devices like network-on-chip \cite{karkar2022thermal,fuschini2018ray,schafer2018chip}, wearable devices for on-body networks \cite{berkelmann2021antenna,turner2012novel} , in which the path loss is not the top priority, the 3D-printed platform provides a fast and feasible solution. Nevertheless, we believe that the proposed reconfigurable surface wave platform will be able to support more applications when new low-loss 3D-print resins are available with advancing material technologies.

\begin{figure}[]
\centering
\includegraphics[width=0.9\columnwidth]{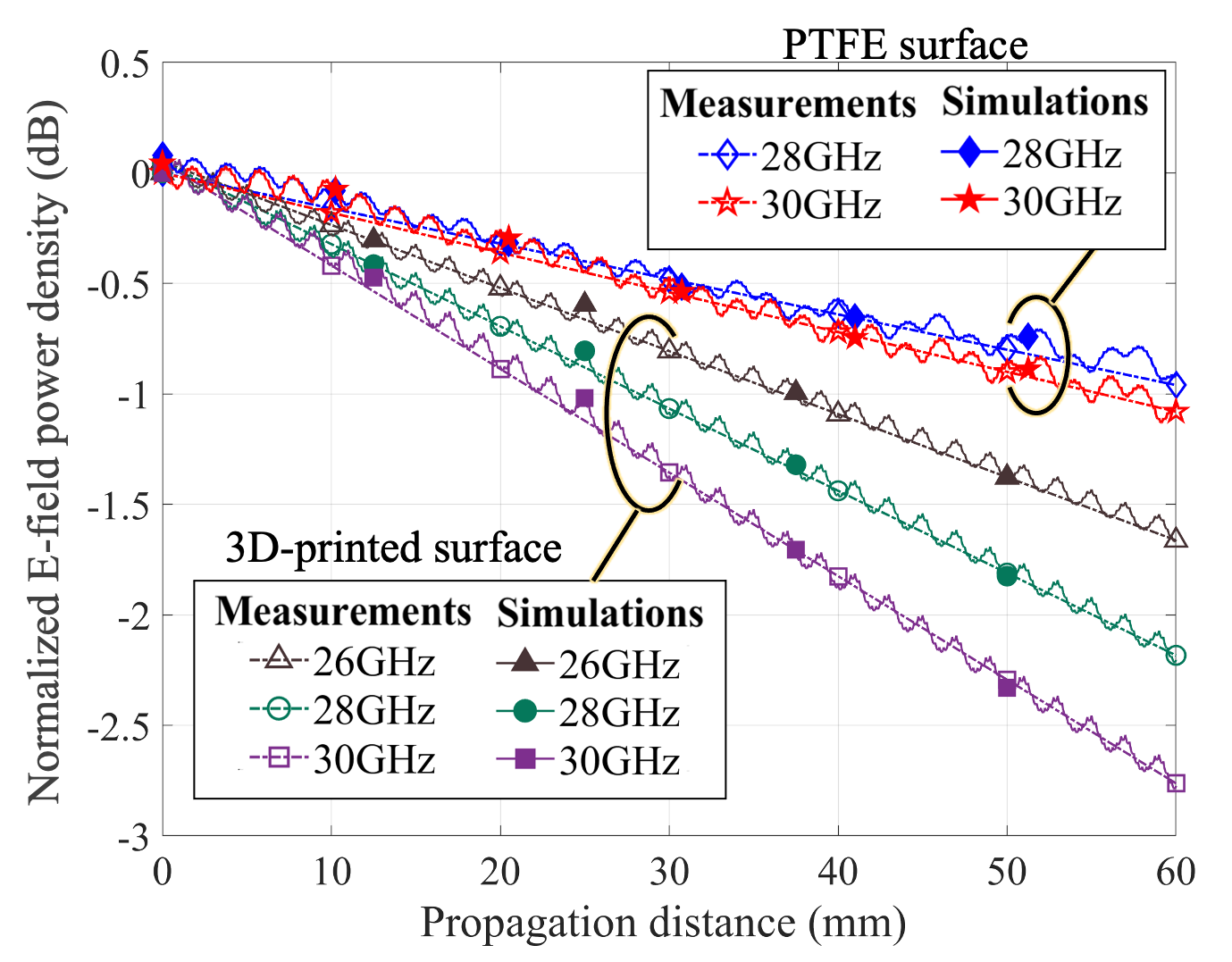}
\vspace{-0.5cm}
\caption{Normalized E-field power density on a straight pathway against the propagation distance at different frequency.}\label{single_layer_result2}
\end{figure}

\subsection{Path Loss Study} 
As depicted in Fig.~\ref{Basic_3D_prototype}, the position of Transducer $ 2 $ can be adjusted by shifting the optical linear stage in the $ z $-direction for measuring the ${\rm S}_{21}$ at different propagation distances from Transducer $ 1 $. Fig.~\ref{single_layer_result1} presents both the measured and simulated ${\rm S}_{21}$ results for the straight pathways using common PTFE and 3D-print resin, with a pathway width $w_c=10\,{\rm mm}$, created by single-layer silver walls, over a propagation distance $d$ ranging from $90\,{\rm mm}$ to $150\,{\rm mm}$ with a sampling interval of $10\,{\rm mm}$.  The results in Fig.~\ref{single_layer_result1} show that the ${\rm S}_{21}$ of both common PTFE and 3D-printed surfaces decay along with the frequency in the range from $23\,{\rm GHz}$ to $30\,{\rm GHz}$, i.e. the proposed 5G millimeter-wave frequency band. There are mainly two reasons, first, the surface impedance will deviate from its optimal value for the best surface wave excitation efficiency \cite{wan2019simulation} at different frequency.  It can be observed that the surface impedance is less susceptible to frequency in the case of common PTFE, as its lower dielectric constant.  Such feature is further evidenced by comparing the 3-dB half-power frequency bandwidth of the two cases.  The 3-dB half-power frequency bandwidth of common PTFE is much wider than the $3.5\,{\rm GHz}$ ($23.7$ to $27.2\,{\rm GHz}$) provided by the 3D-printed surface. Second, the dielectric loss tangent increases with frequency, leading to lower ${\rm S}_{21}$ at higher frequency band. With reference to the gradient of the lines, we can also see that the loss tangent of the 3D-printed surface is more frequency sensitive showing a faster decay with increasing frequency.  The optimal frequency occurs around $25\,{\rm GHz}$ to around $26\,{\rm GHz}$, which corresponds to the peak of the ${\rm S}_{21}$ curve.  In addition, we can observe that the measurement and simulation results are generally consistent. However, the experimental ${\rm S}_{21}$ values are lower than those in simulations below around $25\,{\rm GHz}$. This is possibly due to the cut-off frequency of the two coaxial-to-waveguide adapters as shown in TABLE \ref{Table1} that are connected to the transducers in the measurements \cite{wr28}, while in the simulations, waveguide ports are directly connected to the transducers. Furthermore, we can see that this set of ${\rm S}_{21}$ curves gradually drops as the distance increases across the frequency band.

At $28\,{\rm GHz}$, the ${\rm S}_{21}$ value of the PTFE-platform steadily decreases from $-9.22\,{\rm dB}$ at a propagation distance of $90\,{\rm mm}$ to $-10.09\,{\rm dB}$ at $150\,{\rm mm}$, with a total attenuation of $0.87\,{\rm dB}$ over a distance of $60\,{\rm mm}$, i.e., $14.5\,{\rm dB}$ per metre. The normalized results (assuming the initial reference point at $d=150\,{\rm mm}$) are plotted in Fig.~\ref{single_layer_result2}, where we compare the measured results and CST simulation. We have also included the results at $30\,{\rm GHz}$ and the corresponding results for 3D-printed resin platform at $26\,{\rm GHz}$, $28\,{\rm GHz}$ and $30\,{\rm GHz}$ for reference. Again, the measurement and simulation results agree very well with negligible discrepancy.

\begin{figure}[]
\centering
\includegraphics[width=0.9\columnwidth]{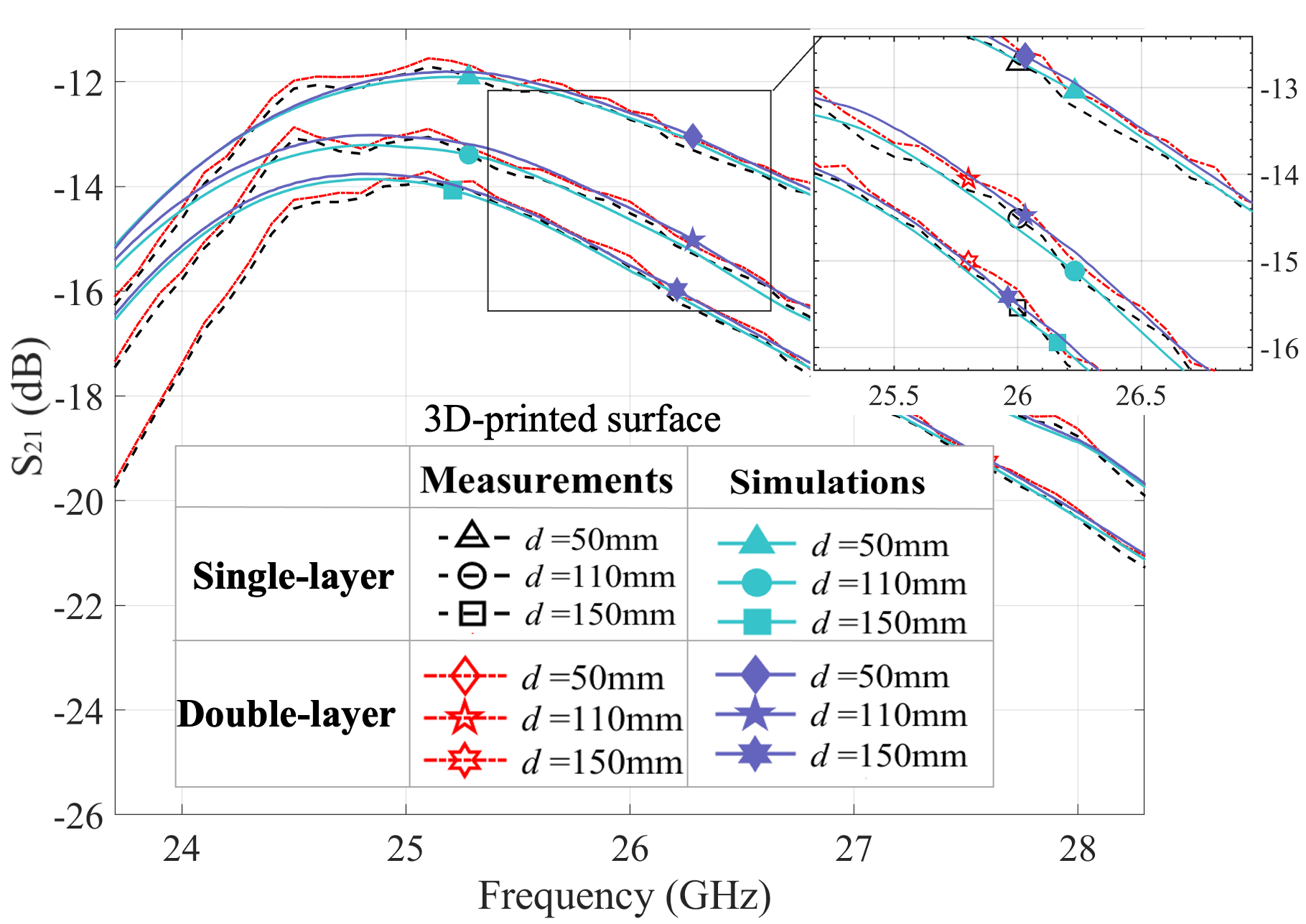}
\vspace{-0.2cm}
\caption{The ${\rm S}_{21}$ results in a straight pathway formed by single-layer and double-layer silver walls in the measurements and simulations with a propagation distance $d=50\,{\rm mm}$, $110\,{\rm mm}$ and $150\,{\rm mm}$.}\label{double_layer_result1}
\vspace{-0.2cm}
\end{figure}

\begin{figure}[]
\centering
\includegraphics[width=0.9\columnwidth]{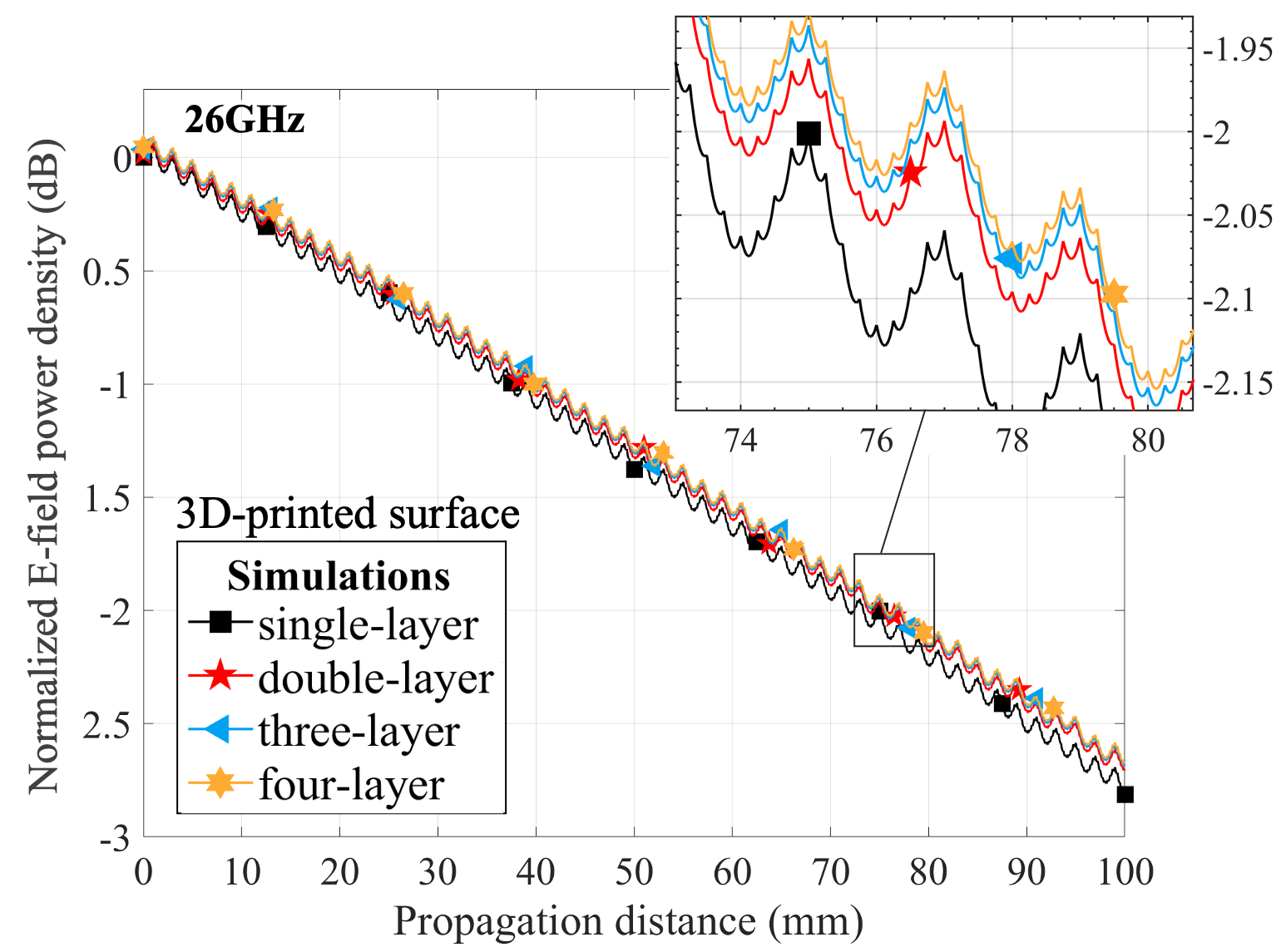}
\vspace{-0.5cm}
\caption{The normalized E-field power in a straight pathway against the distance after normalization for the $(a)$ measurements and simulations containing single-layer, double-layer and $(b)$ multi-layer walls results at $26\,{\rm GHz}$.}\label{double_layer_result2}
\vspace{-0.5cm}
\end{figure}

\subsection{Multi-layer Walls} 
In this sub-section, we investigate the impact of using more layers of metal walls to create the pathway. We first consider the double-layer metal wall setup on the 3D-printed resin platform as depicted in Fig.~\ref{Basic_3D_prototype}$(c)$, and provide both experimental and simulation results of ${\rm S}_{21}$ at several distances for different frequency in Fig.~\ref{double_layer_result1}. It can be observed that the general trend of ${\rm S}_{21}$ for the double-layer and single-layer is very similar, except for a slight increase of approximately $0.1\,{\rm dB}$ in ${\rm S}_{21}$ for the double-layer case. This indicates that the double-layer performs slightly better than the single-layer in confining the signal within the pathway, as the double-layer enhances the ability of the propagation boundary to block signal leakage out of the pathway. Fig.~\ref{double_layer_result2} illustrates the E-field power decay against the distance where the simulation results of three- and four-layer metal walls are also included. It can be seen that the signal improvement in the case of three- and four-layer metal walls is less significant, being only less than $0.05\,{\rm dB}$. Thus, double-layer metal walls appear to be sufficient to create an isolated pathway and guide the surface wave when necessary. In general, the single-layer metal wall structure is adequate to create the reconfigurable surface pathway, as demonstrated in the results above.

\begin{figure}[]
\centering
\includegraphics[width=0.9\columnwidth]{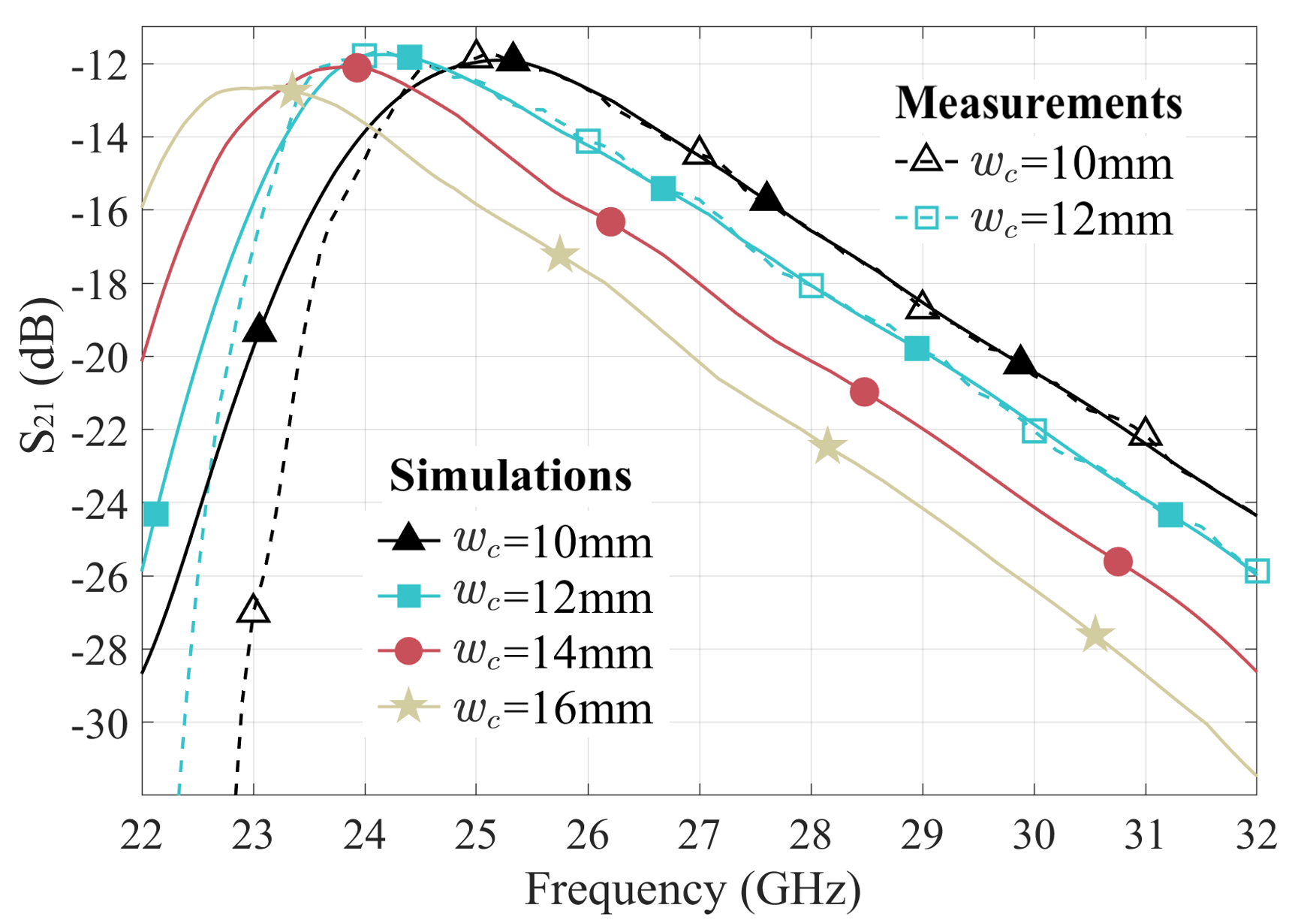}
\vspace{-0.2cm}
\caption{${\rm S}_{21}$ measurements and simulation results in different pathway widths of $10\,{\rm mm}$, $12\,{\rm mm}$, $14\,{\rm mm}$ and $16\,{\rm mm}$ with a propagation distance $d=50\,{\rm mm}$.}\label{Pathway_width1}
\vspace{-0.5cm}
\end{figure}

\subsection{Channel Width and Operating Frequency} 
As depicted in Fig.~\ref{Basic_3D_prototype}$(d)$, the pathway width $w_c$ is increased from $10\,{\rm mm}$ to $12\,{\rm mm}$ defined by two single-layer metal walls in the 3D-print resin platform. Fig.~\ref{Pathway_width1} illustrates the measured and simulated ${\rm S}_{21}$, and the simulated results for $w_c=14\,{\rm mm}$ and $16\,{\rm mm}$ with a fixed propagation distance of $d=50\,{\rm mm}$. The results indicate that the ${\rm S}_{21}$ curves shift to a lower frequency as the pathway width increases, and the optimal operating frequency of the surface changes from $25\,{\rm GHz}$ at $10\,{\rm mm}$ to $24.2\,{\rm GHz}$ at $12\,{\rm mm}$, then $23.7\,{\rm GHz}$ at $14\,{\rm mm}$ and $23\,{\rm GHz}$ at $16\,{\rm mm}$ respectively. Therefore, the pathway width and wavelength of the signal should be positively correlated. We can also foresee the frequency selectivity properties of the pathway width, and it is possible to filter and split the signal at different carrier frequencies by changing the pathway width on this reconfigurable surface. Additionally, we can observe that ${\rm S}_{21}$ decreases slightly with the increase of pathway width, which may be caused by the signal leakage at the transducer to the surface interface. It means that the width of the pathway should preferably be closer to the width of the transducer, which is $w_a=7.1\,{\rm mm}$ here, to minimize leakage.

\subsection{Reconfigurable Pathway} 

\begin{figure}[] 
\vspace{-0.5cm}
\centering 
\includegraphics[width=0.9\columnwidth]{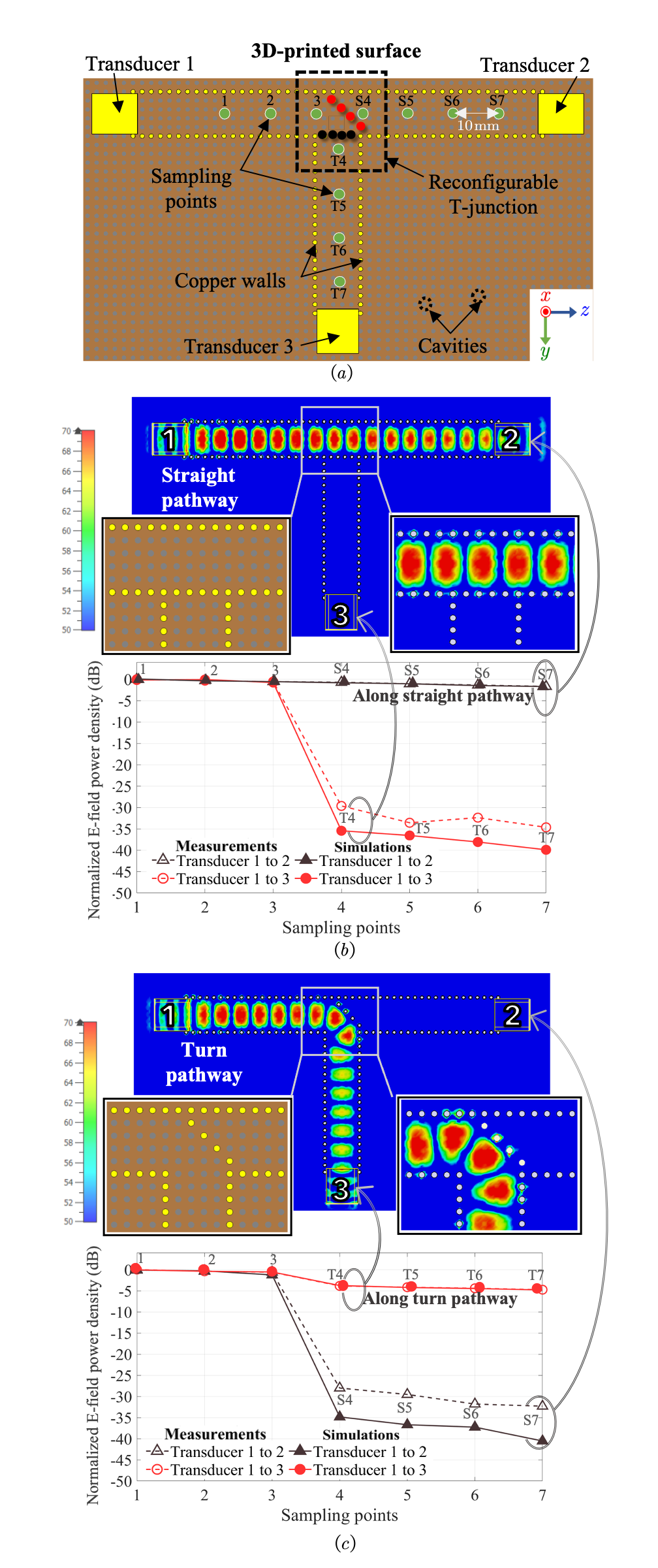} 
\vspace{-0.5cm}
\caption{Illustration of $(a)$ a T-junction reconfigurable surface structure for guiding the surface wave propagation direction along a $(b)$ straight pathway from Transducer $ 1 $ to $ 2  $ or a $(c)$ $90^{\circ}$-bend pathway from Transducer $ 1 $ to $ 3  $ by adjusting the shape of the junction in top view.}\label{Reconfigurable_path1} 
\end{figure}

Fig.~\ref{Reconfigurable_path1}$(a)$ depicts a T-junction reconfigurable surface featuring an adjustable junction in which the positions of metal pins can be altered to switch the surface wave propagation direction between a straight and a $ 90^{\circ} $-bend pathway. In the implementation, the switchable pathway can be achieved by the flow of fluid metal in the cavities, as described in Section II. To demonstrate the concept, we use copper pins with the same radius of the circular cavities to evaluate the surface performance conveniently. In the simulations, eleven E-field sampling probes marked in green shown in Fig.~\ref{Reconfigurable_path1}$(a)$, with an interval of $10\,{\rm mm}$, have been added along the center line of the straight and $ 90^{\circ} $-bend pathways in the simulation models. When copper pins are inserted into the cavities, marked in black, as shown in Fig.~\ref{Reconfigurable_path1}$(a)$ of the T-junction, the surface wave propagates from Transducer $ 1 $ to $ 2 $ along a straight pathway and is blocked from propagating to Transducer $ 3 $, as shown in Fig.~\ref{Reconfigurable_path1}$(b)$. On the other hand, when the pins are re-positioned along a $45^{\circ}$ line (marked in red in Fig.~\ref{Reconfigurable_path1}$(a)$) to direct the signal at the T-junction, the surface wave can be guided from Transducer $1$ to $3$ through a $90^{\circ}$ corner, as shown in Fig.~\ref{Reconfigurable_path1}$(c)$. From the E-field curve, we can observe that there is a difference of more than $30\,{\rm dB}$ between the desired and undesired pathways. The results indicate that the reconfigurable surface can effectively direct most of the surface wave to the desired receiver along an adaptable pathway and reduce the power in other undesirable directions through dynamic pathway selection. Note that the measurement results are sampled by shifting the stages standing the transducer to the probe locations. This T-junction is considered as a proof-of-concept, and the fluid metal pins on the surface can create multiple junctions to change the surface wave propagation directions or bypass obstacles.

\begin{figure}[] 
\centering 
\includegraphics[width=0.85\columnwidth]{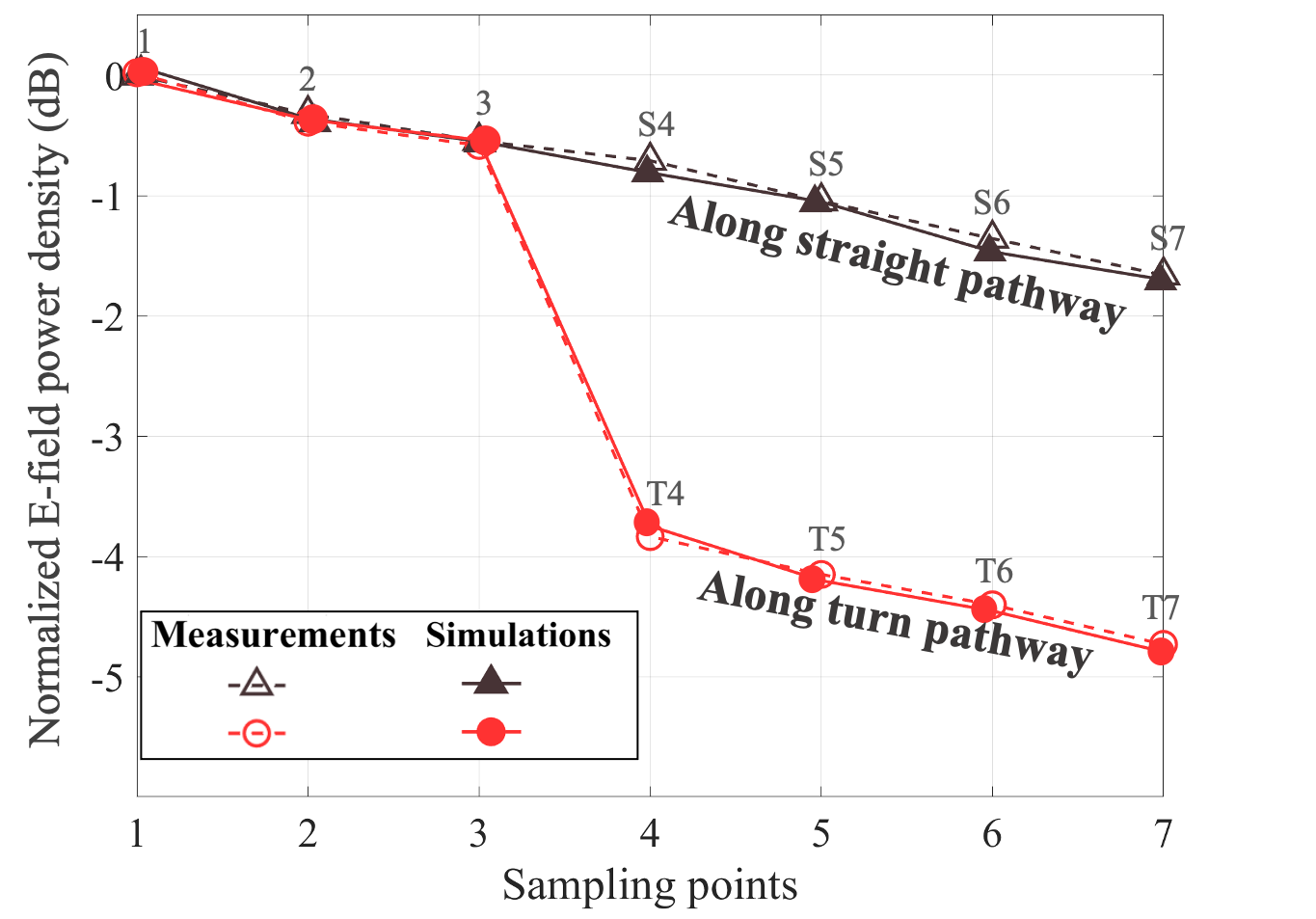} 
\vspace{-0.3cm}
\caption{The comparison of normalized E-field power decay along the straight and $90^{\circ} $-bend pathway in measurement and simulation results.}\label{Reconfigurable_path2} 
\end{figure}

Furthermore, we can directly compare the ${\rm S}_{21}$ results of the straight and $90^{\circ}$-bend pathways using the results in Fig.~\ref{Reconfigurable_path2}. The results reveal that an almost $3.2\,{\rm dB}$ additional loss caused by a single turn can be observed in the $90^{\circ}$-bend pathway, which should result from the signal reflection at the turn. This not only forms a standing wave to increase signal fluctuations but also weakens the signal in the pathway. Additionally, the loss at the corners can be moderately reduced by adjusting the distribution of metal pins to change the corner shape, which will be discussed in the next sub-section.

\subsection{Corner Optimization}  

\begin{figure}[] 
\centering 
\vspace{-0.5cm}
\includegraphics[width=0.85\columnwidth]{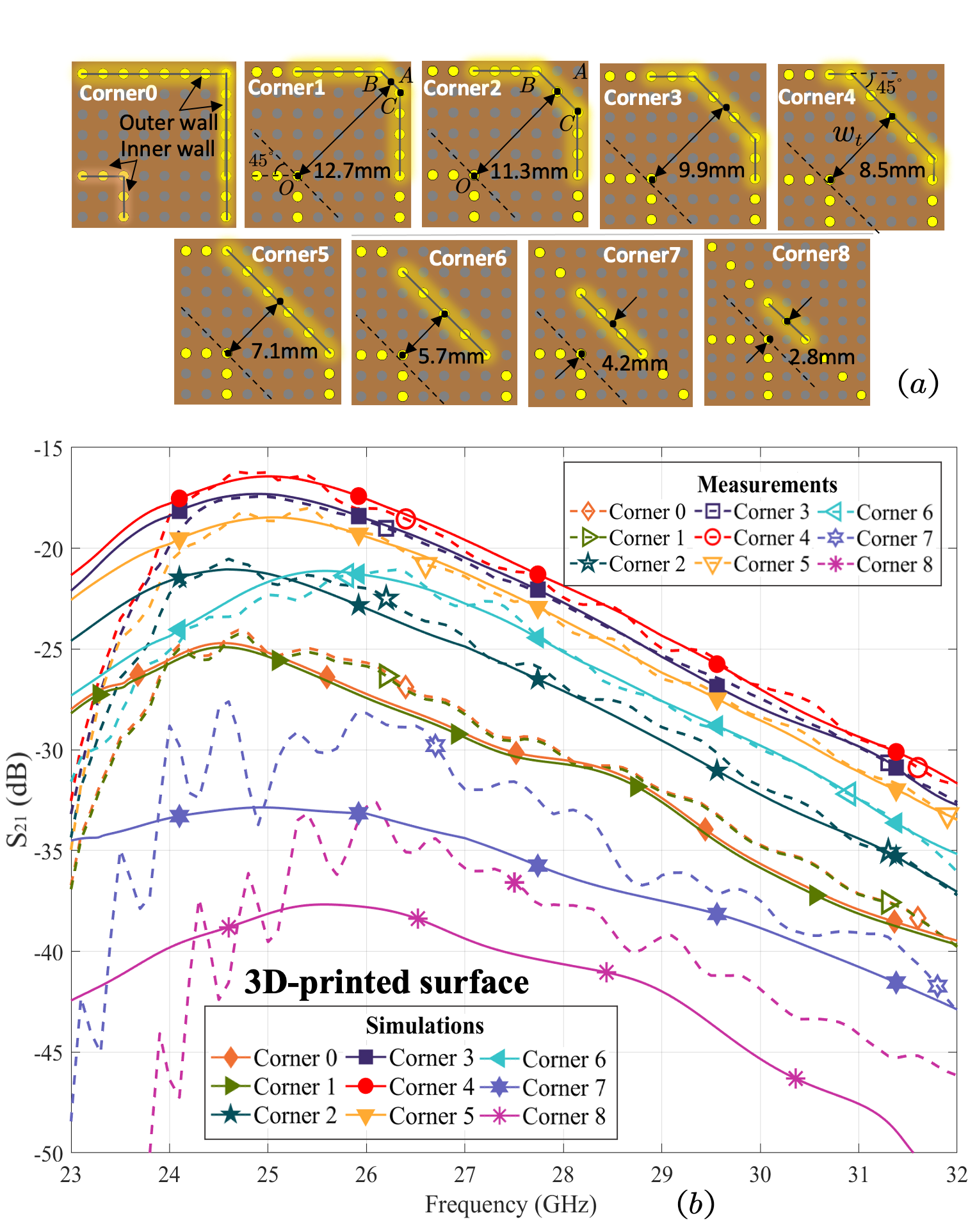} 
\caption{$(a)$ A set of corners with different shapes and corner widths $w_t$ by adjusting the outer metal wall location from Corner 0 to 8 and their $(b)$ ${\rm S}_{21}$ measurement and simulation results.}\label{Corner_shape1} 
\vspace{-0.5cm}
\end{figure}

To study the losses associated with different corner configurations, we replace the T-junction in Fig.~\ref{Reconfigurable_path1}$(a)$ with corners of varying shapes in a $90^{\circ}$-bend pathway for measurement, as illustrated in Fig.~\ref{Corner_shape1}$(a)$. Corner $ 0 $ is used as a standard right angle turn reference, consisting of inner and outer metal walls both at $90^{\circ}$. Corner $ 1 $ is similar to Corner $ 0 $, but with the outer vertex pin removed and the shapes of the pathway in Corner $ 0 $ and Corner $ 1 $ being the same, see the ${\rm S}_{21}$ results in Fig.~\ref{Corner_shape1}$(b)$. Here, we have defined the distance from the inner vertex pin ${\rm O}$ to the center point ${\rm A}$ of the outer wall ${\rm BC}$ as the corner width $w_t$. The corner width decreases as the outer wall ${\rm BC}$ gradually moves towards the inner wall from Corner $ 1 $ to $ 8 $.

\begin{table}[]
\caption{Optimal frequencies of corners with different corner widths}\label{Table3}
\vspace{-0.5mm}
\centering 
\setlength{\tabcolsep}{1.6mm}{}
\begin{tabular}{c|cccccccc}
\hline
Corner                                                                & $1(0)$ & $2$ & $3$ & $4$ & $5$ & $6$ & $7$ & $8$\\ \hline
\begin{tabular}[c]{@{}c@{}}Corner \\ width, $w_t$\\ (mm)\end{tabular} & $12.7$ & $11.3$ & $9.9$ & $8.5$ & $7.1$ & $5.7$ & $4.2$ & $2.8$\\ \hline
\begin{tabular}[c]{@{}c@{}}Optimal\\ frequency, $f$\\ (GHz)\end{tabular} & $24.4$ & $24.5$ & $24.9$ & $25.0$ & $25.1$ & $25.6$ & $26$ & $26.1$\\ \hline
\end{tabular}
\end{table}

The ${\rm S}_{21}$ results for these corners are shown in Fig.~\ref{Corner_shape1}$(b)$. The best shape is Corner $ 4  $ ($w_t=8.5\,{\rm mm}$) with ${\rm S}_{21}$ of $-16.2\,{\rm dB}$ at $25\,{\rm GHz}$, which can be considered as the optimal operating frequency. This is followed by Corner $ 3 $ ($w_t=9.9\,{\rm mm} $) with ${\rm S}_{21}$ of $-17.4\,{\rm dB}$ and Corner $ 5 $ ($w_t=7.1\,{\rm mm}$) with ${\rm S}_{21}$ of $-18.6\,{\rm dB}$. All their corner widths $w_t$ are close to the pathway width $w_c=10\,{\rm mm}$. We also observe that the optimal operating frequency is $24.5\,{\rm GHz}$ in Corner $ 2 $ ($w_t=11.3\,{\rm mm}$) and $25.6\,{\rm GHz}$ in Corner $ 6 $ ($w_t=5.7\,{\rm mm}$), respectively. This discrepancy further illustrates the frequency selection characteristics that can be achieved by controlling the difference in pathway width. Therefore, signal separation based on frequency may be possible in different propagation directions at a junction or corner, which can be considered as a signal filter. For Corner $ 7 $ ($w_t=4.2\,{\rm mm}$) with ${\rm S}_{21}$ of $-30\,{\rm dB}$ and Corner $ 8 $ ($w_t=2.8\,{\rm mm}$) with ${\rm S}_{21}$ of $-33\,{\rm dB}$, the attenuation looks much more significant. This is because the shrinking $45^{\circ} $ outer wall results in more signal reflection and blockage due to the narrow corner width. Note that in Corner $ 7 $ and $ 8 $, the E-field power is too weak and similar to the level outside the pathway, leading to a larger discrepancy between measurement and simulation.

\begin{figure}[] 
\centering 
\includegraphics[width=0.85\columnwidth]{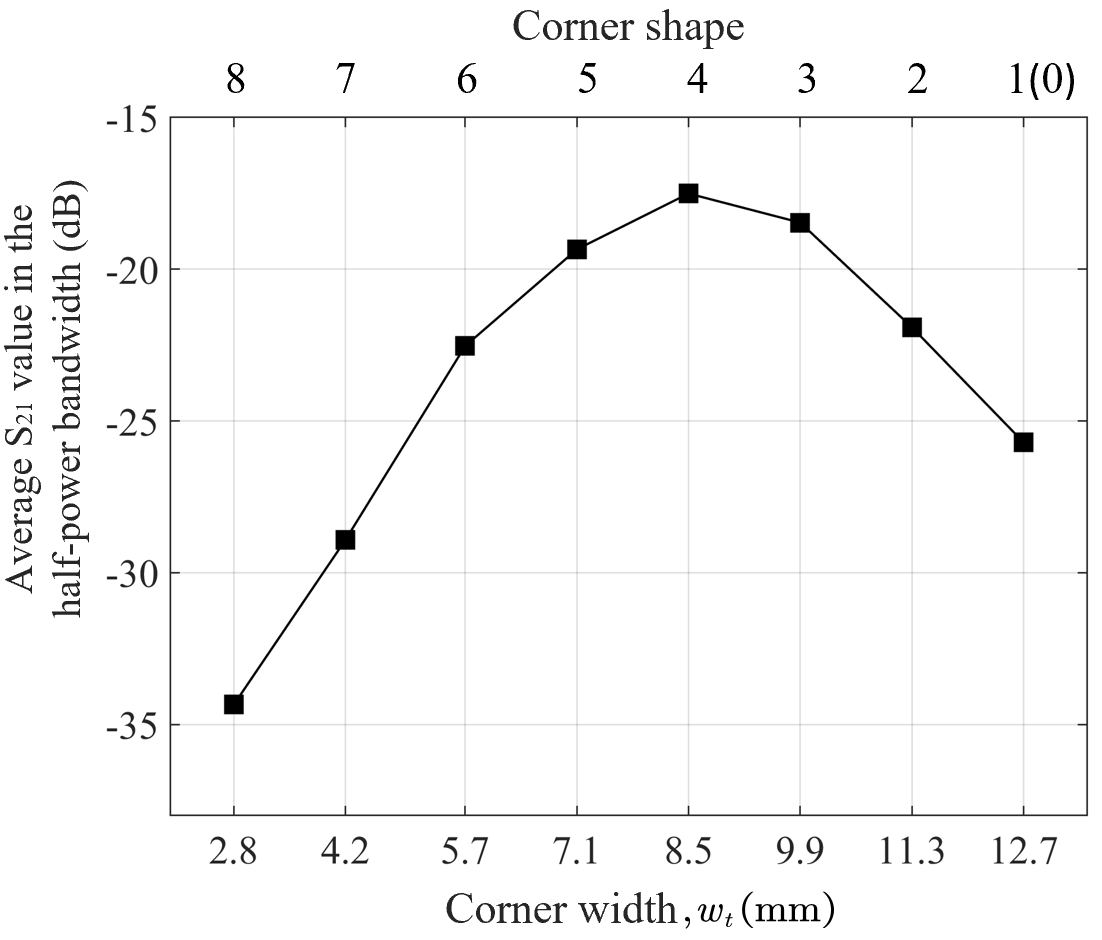} 
\vspace{-0.5cm}
\caption{The relationship between corner width $w_t$ and the average ${\rm S}_{21}$ value in the half-power bandwidth.}\label{Corner_shape2} 
\vspace{-0.5cm}
\end{figure} 

The optimal frequencies of corners with different corner widths are listed in TABLE \ref{Table2}. Fig.~\ref{Corner_shape2} shows the relationship between the corner width and the average ${\rm S}_{21}$ value in the half-power bandwidth based on their respective optimal frequency. We can see that the ${\rm S}_{21}$ value of Corner $ 4 $ with $w_t=8.5\,{\rm mm}$ is higher than that of Corner $ 1 $ with $w_t=12.7\,{\rm mm}$ by over $15\,{\rm dB}$, indicating that the shape of the corner plays a significant role in guiding the surface wave around a $90^{\circ}$ turn. Moreover, Corner $ 4 $ performs slightly better than Corner $ 3 $ with $w_t=9.9\,{\rm mm}$, which is closer to the pathway width $w_c=10\,{\rm mm}$. This suggests that moderately decreasing the corner width, such as in Corner $ 4 $, will result in less loss under the premise of $w_t$ approaching $w_c$ in practice.

\section{Conclusion}
This paper presented a novel porous reconfigurable surface wave platform that takes advantage of fluid metal to dynamically create a customized pathway for guiding surface wave propagation. The concept is driven by the emerging flexible fluid metal micro-fluids technology that is programmable. We analyzed the effect of path loss caused by dielectric materials with different loss tangents, pathway widths, and multi-layer metal wall geometry structures on surface wave propagation using a 3D-print resin surface. We also provided numerical results showing that the proposed reconfigurable surface could outperform traditional space waves and coaxial cables in long propagation distances if the dielectric loss tangent of surface is small enough. Moreover, we presented a switchable pathway process of this reconfigurable surface via a T-junction by re-arranging the formation of metal pins. We demonstrated that surface wave propagation from a straight to a $90^{\circ}$-bend pathway is feasible. Additionally, the results illustrated that properly designing corner width could effectively reduce insertion loss when the surface wave passed through a $90^{\circ}$-corner. In future, we will study the potential in dividing the surface waves into propagation directions. In summary, our work showed the potential of surface wave propagation on a low-loss reconfigurable surface for future communication systems.



\end{document}